# 4D Optically Reconfigurable Volumetric Metamaterials

*Dmitry Dobrykh, Anna Mikhailovskaya, Pavel Ginzburg and Dmitry Filonov[*]*

Dmitry Dobrykh, Anna Mikhailovskaya, Prof. Pavel Ginzburg
School of Electrical Engineering, Tel Aviv University, Tel Aviv 69978, Israel

Dmitry Dobrykh, Anna Mikhailovskaya
Department of Nanophotonics and Metamaterials, ITMO University, St. Petersburg 197101, Russia

Dmitry Filonov
Center for Photonics and 2D Materials, Moscow Institute of Physics and Technology, Dolgoprudny 141700, Russia
dimfilonov@gmail.com

**Abstract -** Metamaterials are artificially created media, which allow introducing additional degrees of freedom into electromagnetic design by controlling constitutive material parameters. Reconfigurable time-dependent metamaterials can further enlarge those capabilities by introducing a temporal variable as an additional controllable parameter. Here we demonstrate a first-of-its-kind reconfigurable *volumetric* metamaterial-based scatterer, wherein the electromagnetic properties are controlled dynamically with light. In particular, hybridized resonances in arrays of split ring resonators give rise to a collective mode that presents properties of artificial high-frequency magnetism for centimeter waves. Resonant behavior of each individual ring is controlled with a photocurrent, which allows the fast tuning of macroscopic effective permeability. Thus, the artificial gigahertz magnon resonant excitation within a subwavelength spherical scatterer is governed by light intensity. Four-dimensional control over electromagnetic scattering in both space and time opens new venues for modern applications, including wireless communications and automotive radars to name just a few.

**Introduction**

Metamaterials are artificially created media in which the electromagnetic properties can be controlled by subwavelength structuring of their unit cells.[1,2] As a result, quite peculiar values of material susceptibilities can be obtained. Widely employed metamaterial designs are based on arrays of compact resonators, which individual response hybridize and give rise to collective modes. Many different approaches, based on variations of loops (magnetic resonators), wires (electric), and their combinations have been developed and experimentally demonstrated.[3-9] The metamaterial approach has been found to be especially successful under conditions where incident illumination wavelength is relatively long. For example, GHz metamaterials can be fabricated by employing standard printed board lithographic techniques. Thus, the field of radio frequency (RF) antennas has benefited from introducing material degrees of freedom as an additional optimization parameter, as was successfully demonstrated in many recent studies.[10-12] It is worth noting, however, that the vast majority of metamaterial-based devices are static, i.e., their electromagnetic properties cannot be controlled after fabrication. This low flexibility can be a significant disadvantage in many cases. For example, a typical invisibility cloaking device has a very narrow operational bandwidth, which is a direct consequence of the resonant nature of the negative index metamaterial that was employed in the celebrated demonstration.[13] However, for application in radar invisibility, the negative index type of countermeasure fails against any basic frequency hopping system, which utilizes a sufficient bandwidth.[14] Hence, the ability of dynamic control over metamaterial parameters is a highly-demanded functionality and has a very broad range of practical applicability, not necessarily limited to radar scenarios.

Control over effective material properties can be achieved in many ways, including, for example, mechanical deformation and electronic control that are among the preferable and already proven strategies.[15,16] The latter approach utilizes electrically driven elements where impedance can be controlled by either external voltage or current. Two-dimensional versions of metamaterials (metasurfaces) especially benefited from this approach, and many useful practical devices have been developed.[10] For example, it was shown that laws of refraction can be controlled almost on demand.[17] Furthermore, several beam steering devices, where expensive phase shifting elements were replaced by low cost varactor diodes, were designed and have demonstrated remarkable electromagnetic performance.[18] Other reconfigurable metasurfaces capabilities have been demonstrated, e.g.[19-22]. However, the direct mapping of those techniques from two-dimensional prototypes to volumetric architectures is extremely challenging. The reason is the need for conducting wiring, which must connect each individual resonating element inside the volume to a drive. The result is that an undesirable branched network of conductors is created inside a structure and start affecting, even if not prevailing over its electromagnetic properties.

Here we propose a paradigm solution to this dynamic reconfigurability problem and demonstrate a first-of-its-kind tunable volumetric metamaterial. In particular, our architecture is based on arrays of split ring resonators (SRRs), serving as microscopic magnetic dipoles. Each SRR is loaded with a varactor, which in turn is driven by a photodiode. The tandem is activated with light and does not have any physical connection (wires) to an adjutant unit cell. The light is guided into the volume of the structure by optical fiber bundles that do not scatter centimeter waves significantly (see the inset in **Figure 1 (a)**).

Under uniform illumination, the collective resonance of the structure can be reconfigured with light. In particular, we show that artificial magnon resonance of a metamaterial-based sphere can be controlled with light, and the resulting scattering cross-sections can be tuned dynamically.[23]

The rest of this manuscript is organized as follows: The metamaterial design for achieving negative permittivity is described first and is then followed by optical control over relative material parameters, numerical demonstration of tunable artificial magnon resonance, experimental observation of electro-optical spectral shift in one unit cell, and backscattering resonance peak degeneration in metamaterial for different powers of non-uniform illumination.

**Metamaterial Design**

Metamaterial architecture is based on ordered arrays of SRRs, chemically etched on printed circuit boards (PCB).[2] Each individual resonator acts as a magnetic dipole within vertices of an artificial material crystal. The specific geometry of the deeply subwavelength unit cell appears in Figure 1 (c). Near-field coupling between all unit cells in the artificial crystal, governs the electromagnetic properties of the homogenized structure. Many different homogenization techniques have been developed over the years that allow deriving effective material parameters of quite sophisticated structures, e.g.[24-26] Our unit cell design contains four split rings and interconnecting wires (within a single cell only). This architecture was chosen to reduce the number of light-rectification photodiodes discussed in the next paragraph. It is also noteworthy that the adjutant unit cells do not have direct wire connection with each other; and, as a result, there is no conduction current flow between the cells. Conductive coupling between the metamaterials' unit cell can give rise to difficulties in defining local effective susceptibilities, as shown, for example, in the case wire media.[27-29] Another strategy to prevent high-frequency current flow is to introduce decoupling coils. This approach, however, can significantly complicate the design and necessitate the use of lumped elements with high tolerance in their nominals. This approach is also associated with additional ohmic losses within a structure. Furthermore, the form factor of lumped elements starts affecting electromagnetic scattering from the entire structure, which can result in additional degradation of performance. To extract the effective parameters of our structure, the numerical retrieval in a waveguide geometry was done with the help of CST Microwave Studio, based on the Nicolson–Ross–Weir method.[30] A typical two-port waveguide system, with relevant dimensions of $20 \times 20 \times 60$ mm, determined by the specific frequency range, was used. Complex-valued transmission and reflection coefficients (S-parameters) were calculated for three different orientations, corresponding to the main crystallographic axis of the metamaterial, and the permittivity and permeability tensors were extracted, following the formulation reported at.[4] The following parameters of the structure were chosen after a set of optimizations: radius of SRRs, r = 4 mm; w = 1 mm and C = 680 pF (the static capacitor within each ring). The substrate was taken to be FR4 ($\varepsilon$ = 4.43, tg$\delta$ = 0.025) of 1.5 mm thickness, and the material chosen for the SRR was copper (conductivity of 5.96 × 107 S/m) (Figure 1 (c) for the geometrical definitions).

The resulting effective permittivity and permeability, as a function of frequency, were extracted numerically and are shown in **Figure 2 (a)**. Both these quantities have a strong resonant absorption peak that results in significantly modified effective values of the real parts in its vicinity. The magnetic phenomenon, which is the subject for the optimization, is stronger in the case of sharp resonances. It is worth mentioning that the magnetic and electric phenomena are decoupled in cases when deeply subwavelength scatterers are considered. Hence, electromagnetic properties, related to relative permeability ($\mu$) values, e.g. -2 for achieving magnon resonance in a deeply subwavelength sphere, do not depend on values of a relative permittivity $\varepsilon$ (although a far-field scattering pattern will be affected). A broad range of negative permeability values can be obtained, ensuring a resonant response even in the presence of retardation effects ($\mu \approx -2$ condition is strictly valid only for point-like scatterers and changes with the non-negligible size of a scatterer).

Artificial magnon resonance resembles a celebrated phenomenon of localized plasmon resonances, taking place in noble metal nanoparticles. Duality of Maxwell's equations allows drawing analogies between electric and magnetic phenomena, while the field of metamaterials enables replicating material properties. The polarizability of a small magnetic sphere is given by

$$\alpha_m = 4\pi r^3 \left(\frac{\mu_r - 1}{\mu_r + 2}\right) \quad (1)$$

where $\mu_r$ is the relative permeability of the material, while r is the radius of the sphere. The modified expression, which includes retardation effects and radiation reaction is given by

$$\alpha_m = 4\pi \left(\frac{1}{r^3} \times \frac{\mu_r+2}{\mu_r-1} - i\frac{2}{3}k^3 - \frac{k^2}{r}\right)^{-1} \qquad (2)$$

where $k$ is the free space wavenumber of the incident radiation. Eq. 1 shows that the resonance is obtained when $Re(\mu_r) = -2$ and the quality factor of the peak is governed by the radiation losses (Eq. 2). More detailed this phenomenon is described in work.[23] It should be noted that the material properties are frequency-dependent and, hence, photodiodes allow controlling the resonant condition (Eq. 2) by changing capacitances of varactors.

Tunability properties of the structure will be assessed next. For this purpose, several capacitance nominals of the varactor diodes were chosen to demonstrate the tunability characteristics. Decrease in capacitance (demonstrated in the following experiment with light) shifts the resonance of the structure to higher frequencies, which is quite expected from the lumped circuit theory (Figure 2 (b)). An efficient tuning can be obtained within the frequency band, corresponding to 10% of the carrier frequency. In many practical applications, 10% bandwidth is quite sufficient for a normal performance.

Tolerance in lumped element nominals can be quite important if many of them are used in a structure with a strong resonant response. Furthermore, as it will be shown in the experimental part, the capacitance tuning will be performed with light. It means that non-uniform illumination of photodiodes and different efficiencies of the later elements will result in a pseudo-random spread of capacitances within a unit cell. To estimate the impact of these factors on performance degradation, a statistical model was developed. The parametric retrieval was performed on six unit cells, placed inside the waveguide with dimensions $20 \times 20 \times 60$ mm. Capacitance nominals within each unit cells were taken to be equal since the same photodiode drives the voltage drop on the elements. However, nominals in the different unit cells were taken according to the following statistical model:

$$C_{var} = C_0 - \Delta C_{var} \cdot rand([0,1]),  \quad (3)$$

where $C_0 = 9.2\ pF$ and $\Delta C$ is varied from zero (all the elements are the same without a spread) to 1.2 pF (corresponding to the maximum $\Delta C_{var}$ for the Skyworks SMV 1413 varactors, driven by one PIN photodiode BWP34) and $rand([0,1])$ is a random variable with a uniform distribution between 0 and 1. This particular form was chosen to resemble the experimental scenario, where $C_{var} = C_0$ is the capacitance in the "dark" mode, and its value can only be reduced with the introduction of a photocurrent. For each specific value of $\Delta C_{var}$ $n = 5$ realizations have been considered. Permeability dispersion was extracted for each particular case and then averaged over the number of realizations. **Figure 3** demonstrates the resonance degradation of the permeability as a function of the increased spread in the nominals, $\Delta C_{var}$. It can be seen that around the critical value, $\vartheta = \frac{\Delta C_{var}}{C_0} = 0.13$, the effective permeability does not reach the value of $-2$. Large values of $\vartheta$ lead to almost complete elimination of the effect, and the relative permeability averages to unity. This effect will be important for analyzing the experimental data, reported in the next section.

**Tunable artificial magnon resonance**

Electromagnetic scattering performance of 3D metamaterial-based sphere is assessed in this section. Numerical analysis was conducted with the help of frequency domain solver, implemented within the CST Microwave Studio software. The final design of the scatterer consisted of 11 PCB layers with printed unit cells, which were assembled to form an approximately spherical shape with a radius of 60 mm. The distance between the layers was h = 9 mm. The geometry and materials of the unit cells were exactly the same as those investigated in the previous section. The structure was illuminated with a plane wave, polarized along the y axis (see **Figure 4 (a)** for the geometric layout). The analysis was made by imposing open boundary conditions. The numerical modeling replicates performances of photodiodes with effective lumped elements. The varactors capasitance change was estimated via a voltage drop, which in turn is poroportional to the light intensity, driving the process. Additional capacitance of photodiodes was taken into account by plugging nominals, declared in a datasheet. The maximal change of 1.2pf was imposed in order to correlate the numerical investigations with the experimental data.

The backward scattering cross section was calculated for the following values of the varactor diodes: $C_{var} = 9.2\ pF$, corresponding to the "dark" regime; $C_{var} = 8\ pF$, corresponding to the uniform "bright" regime; and $\Delta C_{var} = 1.2\ pF$, corresponding to the inhomogeneous illumination case. In the ideal case, the observed resonance shift was around 10% with respect to the carrier frequency (Figure 4 (b)). For an inhomogeneous illumination, however, the resonant scattering peak vanishes for the critical value of $\Delta C_{var}$.

Since the structure is strongly anisotropic, it is worth investigating its scattering capabilities in the case of inclined incidence. The key parameter in this case is the projection of magnetic field polarization on the major axis of the structure. Fig. 5 shows the backscattering cross section for different Euler angles. Panel a corresponds to a normal incidence, where k-vector of the wave is parallel to the planes, containing SRRs and field's polarization is rotated. A drop of the scattering peak with increasing the polarization mismatch can be clearly observed. Fig. 5(b) investigates another scenario, where both k-vector and the H-field polarization are rotated. Similar drop is observed. It is worth noting, however, that 45° polarization mismatch case in scenarios on panels (a) and (b) leads to slightly different peak heights, underlining the contribution of anisotropic permittivity contribution.

**Electro-optical tuning**

To check the experimantal feasibility of light-induced tuning capabilities, a single unit cell was constructed first and its resonant properties were investigated. The unit cell was fabricated on FR-4 PCB (dielectric constant of 4.43, a thickness of 1.5 mm) by chemical etching. Lumped elements (Multilayer Ceramic Capacitors 500R07S0R5AV4T and Skyworks SMV 1413 varactors) were soldered to the SRRs' gaps. The capacitance of each varactor diode was controlled by soldering the BWP34 photodiode on the PCB. Complex reflection coefficient ($S_{11}$) of the structure was acquired with the help of a small magnetic probe antenna connected to the transmitting port of a Vector Network Analyzer (VNA E8362B). The probe was placed 5 mm above the unit cell and two regimes of illumination were tested: "dark" with no light source and "bright" with the light source turned on. In the "bright" regime, the photodiode was illuminated with 100 mW red laser pointer ($\lambda = 655$ nm). Experimenal results of a single unit cell tuning are shown in **Figure 6 (a)** that clearly show the efficent resonace shift of 10% with respect to the carrier frequency was achved.

**Electro-optical tuning of the artificial magnon resonance**

One of the main challenges in constructing volumetric tunable metamaterial is to provide a drive to each individual cell. Electrical wiring of a circuit board significantly affects antenna performance by introducing additional coupling channels. Furthermore, the introduction of bunched metal strips influences the scattering characteristics; and, in fact, can even govern the interactions with incident radiation. This problem can be partially solved in reflect arrays configuration, where the wires are hidden behind the ground plane that isolates them from antenna elements. It is important to note, however, that those architectures are planar and cannot be extended to the third dimension – this is the key difference between 2 and 3D. Here we introduce a new concept of electro-optical drive, distributed over an optical network, and implement it experimentally.

The experimental sample was fabricated following the effective medium design, maintaining the same geometric parameters as in the numerical modeling. Overall, the structure has 1080 lumped elements. To provide an efficient scattering cross-section modulation with light, the capacitance of each varactor diode was controlled by the BWP34 photodiodes, exactly as in the case of unit cell discussed earlier. The light was guided to the diodes with the help of wide fiberglass fibers, which were roughened on purpose to out scatter a sufficient optical intensity. Those fibers (a total of 22) can be clearly seen on the photograph in Figure 1. Rectangular wideband horn antenna connected to the transmitting port of a Vector Network Analyzer (VNA E8362B) was used as an excitation source. The scatterer was then located in the far-field of the antenna (2 meters apart). Backward scattering cross section was obtained using the same horn antenna to collect the signal from the sphere. The backward cross section spectra were measured with and without the light illumination, provided in this case by a halogen fiber optic illuminator (Thorlabs OSL2, Thorlabs, Inc.). Relatively small integral power of the red laser, used for investigations of a single cell, limits its use as a device to drive the whole structure. Differential RCS was measured as a difference between backward scattering cross-sections of object in dark and bright regimes, i.e., $\sigma^{dif} = |S_{11}^{dark}| - |S_{11}^{bright}|$. The current realization fails into the category where $\Delta C_{var}$ is above the critical value, corresponding to the degradation of the scattering peak with the light source switched on. This effect is demonstrated experimentally, and the results are shown in Figure 6 (b)—increasing the light intensity leads to the vanishing of the peak—and, hence, to the maximization of the differential RCS (blue line). For moderately low intensity, the light-induced modulation is moderately low, resulting in vanishing differential RCS as a function of frequency. This typical example demonstrates the capability to obtain amplitude modulation of RF signals with light. The light-induced elimination of the scattering peak corresponds to the random nominal spread theory, presented in Fig. 3. Slightly different light intensities, delivered to the unit cells reduce the element's tolerance, which cause to the peak smearing.

**Conclusions**

Volumetric reconfigurable metamaterial was designed and experimentally demonstrated. The concept of tunable RF magnon resonance, controllable with light, have been developed. The architecture is based on arrays of SRRs that serve as microscopic magnetic dipoles, which hybridize to a collective magnetic mode. Each SRR was loaded with a tandem of a varactor and a driving photodiode, which allowed replacing branched conducting wire network with optical fibers that are transparent to RF waves. It was shown that artificial magnon resonance within the metamaterial-based sphere can be controlled by light and scattering cross-sections are affected by it in real time. This demonstration can link to new solutions in the field of wireless communications and RFID technologies, where real-time control over scattering cross-sections is highly demanded. Furthermore, recent advances in additive manufacturing can allow exploring all three geometric degrees of freedom more efficiently and enhance the capabilities with additional temporal control.[31]

**Keywords**

Metamaterials, Artificial magnon resonance, Reconfigurable materials

**Acknowledgments**

The research was supported by the Russian Science Foundation (Project 19-79-10232), ERC StG 'In Motion,' PAZY Foundation, and Israeli Ministry of Science and Technology (project "Integrated 2D & 3D Functional Printing of Batteries with Metamaterials and Antennas".

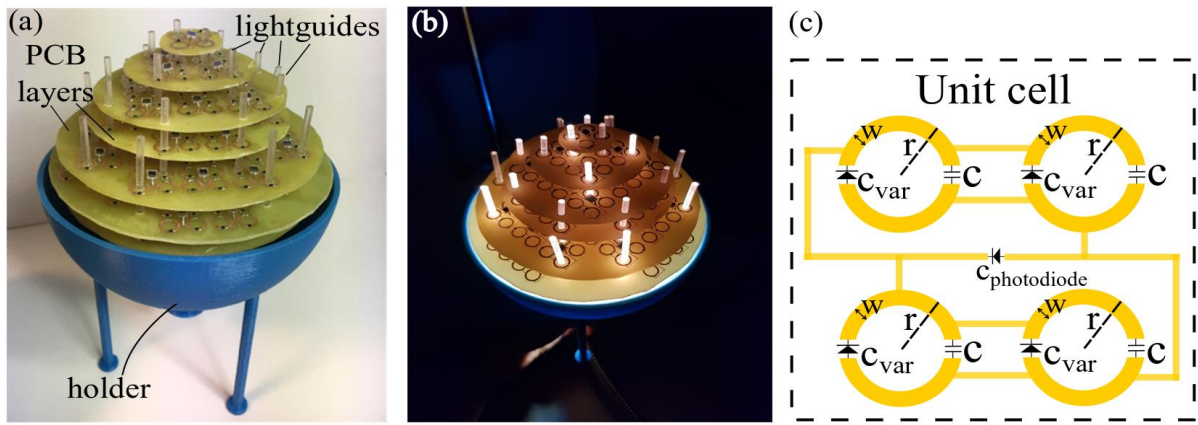

**Figure 1.** Electro-optically tunable artificial magnon resonance in a metamaterial-based sphere (a) Photograph of sphere with a plastic 3D-printed bottom holder. (b) Photograph of sphere with optical drive switched on and the corresponding "bright" regime. (c) Schematics of a unit cell forming the metamaterial

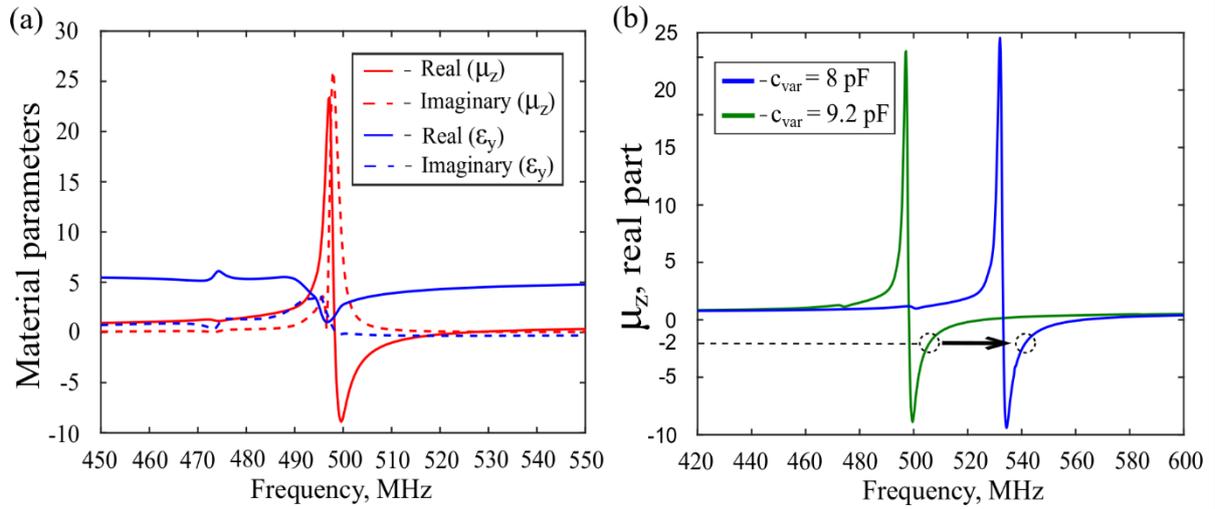

**Figure 2.** (a) Numerically extracted (CST Microwave Studio) effective metamaterial parameters as a function of frequency for the unit cell, presented in Figure 1 (c). Real and imaginary parts of permittivity and permeability are indicated in the legends. (b) Tunability of the real part of numerically extracted effective permeability. Different nominals of the varactor diodes correspond to different light intensities, applied on the photoelement.

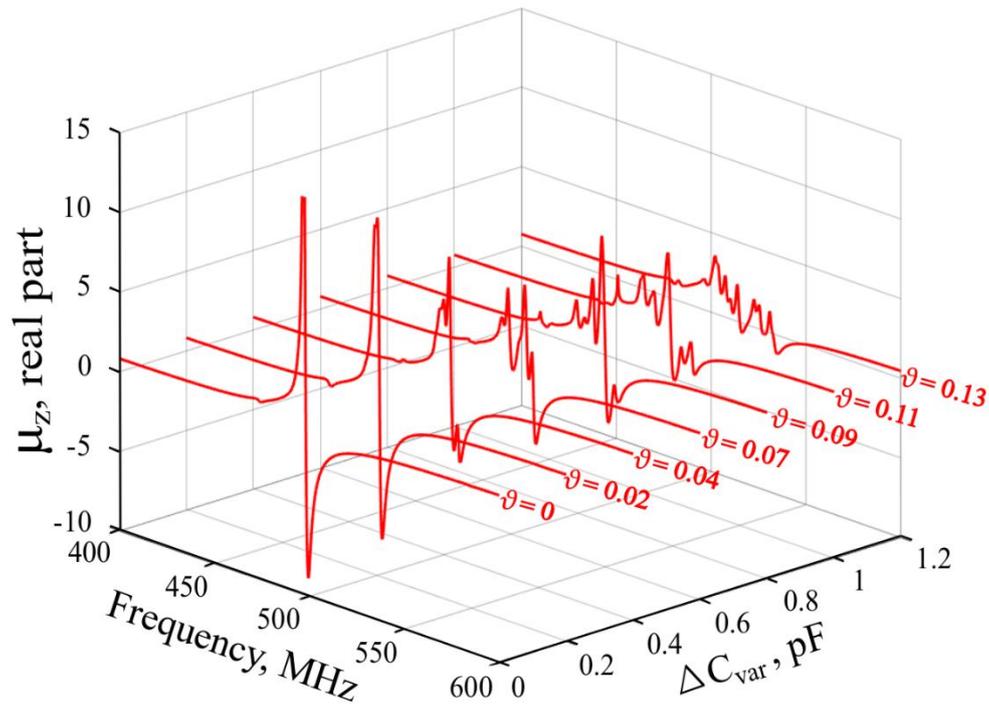

**Figure 3**. Degradation of resonant behavior of the effective permeability with the drop in tolerance of the lumped elements nominals. Real part of the effective permeability as a function of frequency is plotted for different values of the tolerance ($\varDelta C_{var}$ in the figure).

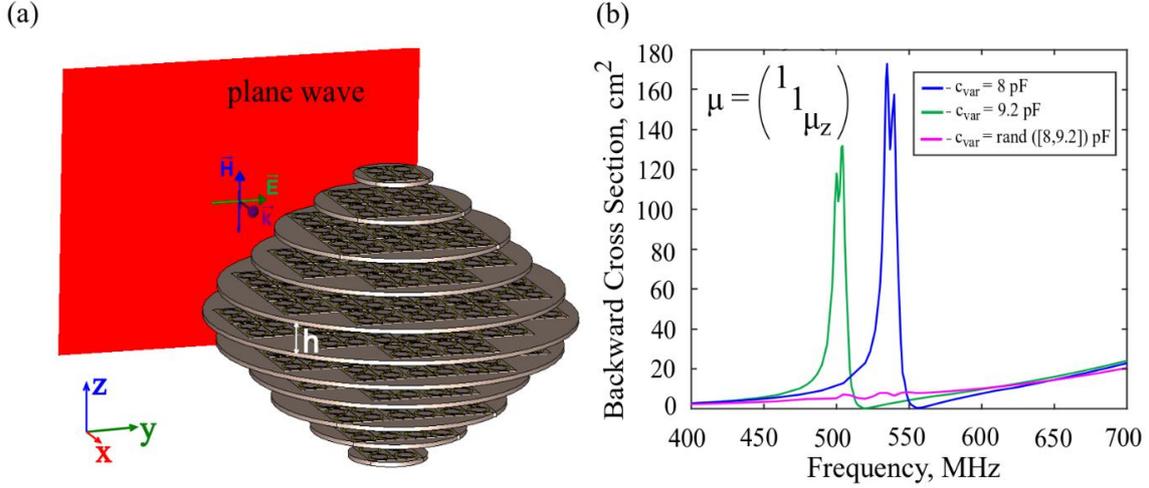

**Figure 4.** Theoretical prediction of the artificial RF magnon resonant tuning with visible light (a) Basic setup for the numerical analysis—a plane wave is incident on the metamaterial-based spherical scatterer (b) Backward scattering cross section spectra of the sphere in "dark" mode—green line, "bright" mode—blue line, and non-uniform illumination ($\vartheta = 0.13$)—magenta line.

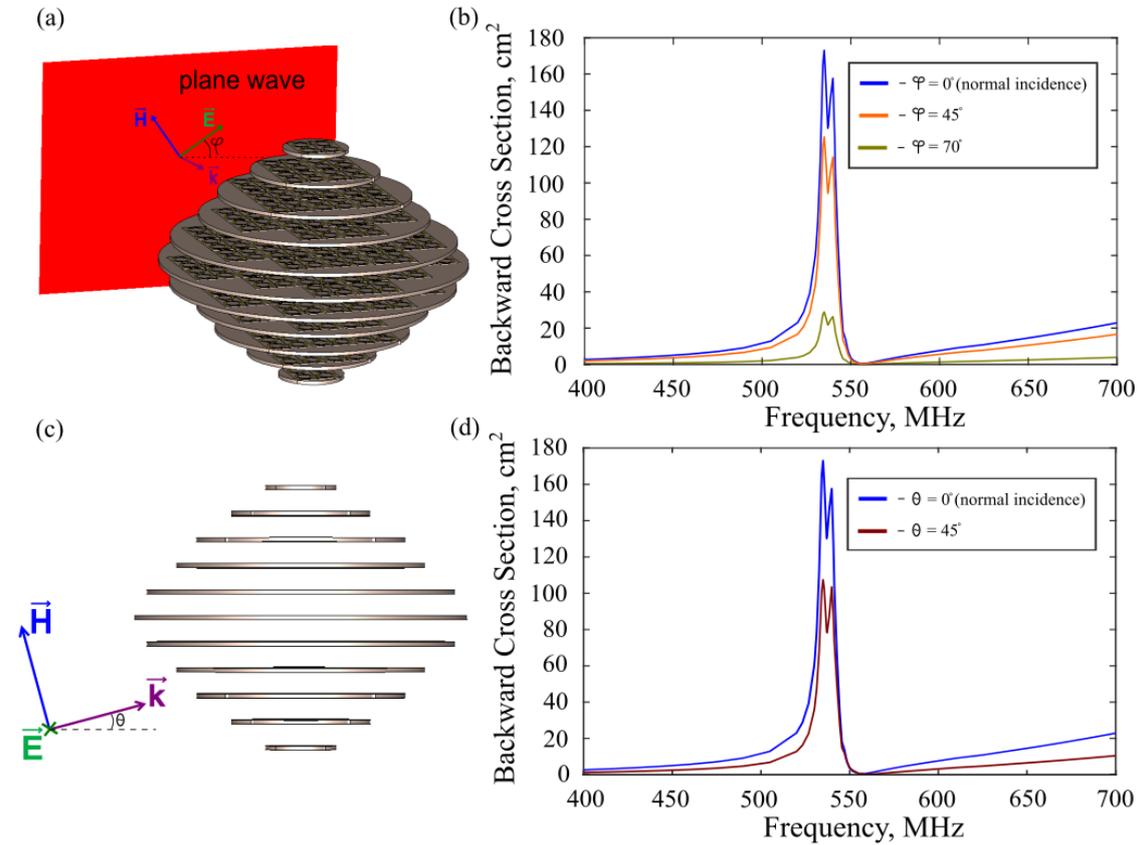

**Figure 5.** Numerical analysis of back scattering spectra for different angle of incidence and polarization. (a) Schematics of the interaction scenario - φ is an angle between electric field and the major axis of the metamaterial, k-vector is in the plane, containing the resonators. (b) Back scattering spectra for different angles (scenario of panel a), indicated in the legend. (c) Schematics of the interaction scenario, $\theta$ is indicated. (d) Back scattering spectra for different angles (scenario of panel c), indicated in the legend.

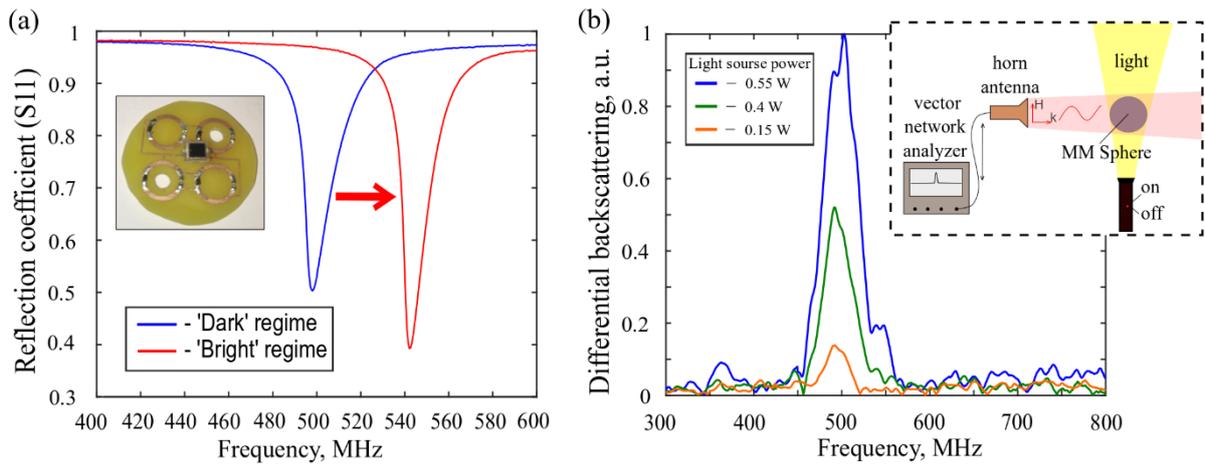

**Figure 6.** (a) Experimental demonstration of optical tunability of a single unit cell. "Dark" regime—no illumination; "bright" regime—the photodiode is illuminated with 100 mW laser (λ =655 nm). S$_{11}$ parameter as a function of frequency is obtained by approaching the structure with a near-field probe (b) Experimental demonstration of optically tuned differential backscattering of an artificial magnon resonance